\documentclass[useAMS,usenatbib]{mn2e}

\usepackage[dvips]{graphicx}

\title[Dense, hot winds from TTSs]{On the source of dense outflows from T Tauri
Stars.\\
and III. Winds driven from the star-disc shear layer.}

\author[G\'omez de Castro and von Rekowski]{Ana I. G\'omez de
Castro$^{1}$\thanks{E-mail:aig@mat.ucm.es} and Brigitta von Rekowski$^{2}$\\
$^{1}$S.D. Astronom\'{\i}a y Geodesia, Fac. de CC. Matem\'aticas, Universidad
Complutense de Madrid, Plaza de Ciencias 3, 28040 Madrid, Spain\\
$^{2}$Institute of Human Genetics, Newcastle University, International Centre
for Life, Newcastle upon Tyne NE1 3BZ, UK}

\begin{document}

\date{Submitted December 7th, 2009}

\pagerange{\pageref{firstpage}--\pageref{lastpage}} \pubyear{2009}

\maketitle

\label{firstpage}

\begin{abstract}
Ultraviolet observations of classical T Tauri Stars (cTTSs) have shown that
there is a hot ($T_{\rm e}\simeq 80,000$~K) and dense ($n_{\rm e}\simeq 10^{10}$
cm$^{-3}$) component associated with the large scale jet. This hot component
is formed very close to the base of the jet providing fundamental information
on the jet formation mechanism. In this series, we have investigated whether this
component can be formed in disc winds, either cool or warm.  
To conclude the series, jet launching from the interface
between the magnetic rotor (the star) and the disc is studied. 
Synthetic profiles are calculated from numerical simulations of outflow launching by 
star-disc interaction. Profiles are calculated for several possible configurations
of the stellar field: dipolar (with surface strengths, $\ B_*$ of 1, 2 and 5~kG) or 
dynamo fed. Also two types of discs, passive or subjected to an $\alpha \Omega$-dynamo,
are considered. These profiles have been used to define the locus of the various
models in the observational diagram: dispersion versus centroid, for the profiles
of the Si~III] line. Bulk motions produce an increasing broadening of the profile 
as the lever arm launching the jet becomes more efficient; predicted 
profiles are however, sensitive to the disc inclination.
Models are compared with observations of the Si~III] lines obtained with the
{\it Hubble Space Telescope}.

In addition, it is shown that the non-stationary nature of star-disc winds
produce a flickering of the profile during quiescence with variations in the
line flux of about 10\%. At outburst,
accretion signatures appear in the profiles together with an enhancement of the
wind, producing the correlation between accretion and outflow as reported from 
RU~Lup, AA~Tau and RW~Aur observations.

\end{abstract}


\section{Introduction}

Star formation is accompanied by heavy mass outflow. Outflows are powered by the
gravitational field (through the centrifugal gear) and by the release of the
magnetic energy built up in the sheared and turbulent protostellar discs.
Since the very early studies of the atmospheres of the T Tauri Stars (TTSs),
it is known that during the  classical T Tauri phase ($\sim 1$~Myr old stars),
TTSs drive powerful and massive cool (T$\sim 10,000$~K) winds whose signature
was first detected in the profiles of neutral and singly ionized species
being the most conspicuous the optical H$\alpha$ line and the Mg~II resonance
multiplet (UV1) in the UV (Giampapa et al 1981, Penston \& Lago 1983,
Finkenzeller \& Mundt 1984, Calvet et al 1985). The realisation that TTSs drive
cool (T$\sim 10,000$~K) jets and bipolar outflows came in the early 80's
(i.e. Schwartz 1983, Mundt et al 1987). Ever since, it
is unclear which fraction of cool wind is associated with the so-called cool
{\it stellar} wind (and which fraction is associated with the bipolar outflow).
For instance, the absorption by this cool wind of the H$_2$ emission from the disc
(produced within $\sim 2$~AU from the star, Herczeg et al 2004), 
clearly indicates that  a significant fraction of the absorbing gas is associated with the large
scale jet.

Though there is a general consensus on bipolar outflows
being driven by the disc-star interaction and the disc wind
(see i.e. Ferreira et al 2006), the nature of TTSs {\it stellar} winds remains open.
Hot winds are expected to be produced by the star since the strong magnetic activity
and the powerful stellar corona (Preibish et al 2004) suggest that a pre-main sequence  
analogue to the solar coronal winds must exist. Unfortunately, this
hot component has not been conclusively detected. This is caused by the complexity of 
the atmospheres and the disc-star interaction region in TTSs that makes difficult
to disentangle the contribution from the photoionized accretion flow and the inner
disc from a theoretically hypothesised hot stellar wind. In 2005, Dupr\'ee
et al, claimed  to have detected, for the first time, this hot wind.
These authors interpreted the absence of blueshifted emission in the
O~VI resonance lines in TW~Hya and T~Tau, as an indication
of the existence of a hot wind with temperatures as high as 300,000~K and
mass-loss rates of $2\times 10^{-11}$M$_{\odot}$yr$^{-1}$ that absorbs
the blue wing of the profile. As pointed out by Johns-Krull \& Herczeg (2007),
the lack of a blueshifted component could just be an indication of the line being formed 
in the accretion flow. Matter infall onto the stellar surface is channelled
by the magnetic field driving to the formation of shocks at the impact points
where the kinetic energy of the infalling gas is finally released. The temperature
reached at the shocks can be as high as some 10$^6$K and the photoionizing
X-ray radiation preionizes the infalling gas column that radiates over a wide 
range of temperatures and tracers (Lamzin 1998).

An independent line of research, the one addressed in this series, was to
investigate whether the base of the jet could be hot enough to radiate
at temperatures as high as 50,000~K-80,000~K, well below the temperature range
of the hot stellar wind but above the fiducial temperatures of optical
jets: 10,000~K-20,000~K.
The motivation for this research came from the discovery of C~III] and
Si~III] semiforbidden emission in the UV spectrum of RY~Tau and RU~Lup
at the same velocity of the optical jet; this emission traces gas at
$\log T_e \sim 4.6-4.8$ (G\'omez de Castro \& Verdugo 2001). Centrifugally
driven flows from magnetised accretion discs are submitted to
pinching stresses $\vec j \times \vec B$ because the toroidal magnetic field
($\vec B$) collimates the  current of  gas along
the disc axis ($\vec j$). Recollimation can drive to the formation of focal surfaces
or shocks on the jet axis that are able to heat the gas to the some
10,000~K traced by the optical forbidden lines of S~II]
(G\'omez de Castro \& Pudritz 1993). However, temperatures as high as
80,000~K cannot be produced in cool disc winds. In the first article
of this series, we examined whether the photoionisation of cool disc winds
by the stellar corona could cause the observed emission (Ferro-Font\'an \& G\'omez
de Castro 2003, hereafter Paper~I). We found that
the propagation of the stellar radiation generates a cocoon of photoionized gas 
around the star. The extent of the photoionized region is small (tenths of au) 
in dense outflows and close to the disc plane; however, it may cover the whole 
wind extent in diffuse winds, e.g. disc winds generated by small accretion rates
($\leq 10^{-9}$~M$_{\odot}$yr$^{-1}$). Photoionisation also modifies the electron 
density in the plasma. The interplay between ambipolar diffusion and the radiation 
field controls the electron temperature of the wind that is kept around 10,000~K, 
well below the temperatures
traced by the UV semiforbidden lines.

Disc winds are a fundamental mechanism for angular momentum transport in protostellar
discs, contributing to the regulation of the accretion rate onto the star.
However, most of the transport occurs in the disc itself through the magnetorotational
instability (Balbus \& Hawley 1991):  weak fields provide a tension force that allows 
two orbiting fluid elements to exchange angular momentum on larger scales than the 
hydrodynamical viscosity scales. The magneto rotational instability
acts like a dynamo, amplifying the field which is lost due to buoyancy, leading 
to a magnetised corona. Flares associated with reconnection events would naturally be produced
leading to the formation of a hot disc corona.
Warmer disc winds, centrifugally launched from the disc corona, were shown to
be able to reproduce the observed line fluxes and  line ratios provided that
the winds are clumpy (with filling factors about 1\%) in G\'omez de Castro
\& Ferro-Font\'an 2005 (hereafter Paper~II), the second article of this series. Warm disc
winds have also been shown to be able to reproduce larger scale jet observations
(see i.e. Vlahakis \& Tsinganos 1999, Ferreira 2004). Warm disc winds provide
an elegant and simple solution to the high jet temperatures observed in the UV
however, they are not adequate to reproduce in full detail the observed physics of the 
line formation region. Long tails of bluewards shifted emission are detected in the Si~III] 
and C~III] plasma tracers, suggesting that the outflow launching
mechanism is more efficient and, at the same time, the outflow is less
collimated than predicted by the simple self-similar warm disc wind theory. 
As shown in Paper~II, the self-similar
solutions that are able to reproduce the observed jet properties (velocity and
mass flow) require that the ratio between the sound speed  and the escape
velocity is 0.43 at the Alfv\'en radius. As a result, the radial expansion of
the outflow - the main source of line broadening - is shifted to distances of
about 10~AU from the star, where plasma cannot contribute to the Si~III] and
C~III] emission because the densities and temperatures are very low and
radiative cooling is dominated by singly ionized species
(see Fig.~3 in G\'omez de Castro \& Verdugo 2007, hereafter GdCV07).
Thus, self-similar models produce winds too collimated at the base to reproduce 
the observed broadening of the Si~III] and C~III] profiles.

The high densities revealed by the Si~III], C~III] and C~IV] line ratios
indicate that this line radiation is produced very close to the star. Henceforth
the radiating plasma must be strongly affected by the disc-star interaction.
In fact, there is evidence that  Si~III] and C~III] radiation can be produced in 
structures very close to the star such as the ionized plasma torus around RW~Aur
(G\'omez de Castro \& Verdugo 2003) or the accreting shell in RY~Tau
(GdCV07).
The physics of the interaction between the stellar magnetic field and the
accretion disc is very rich, and it has been shown extensively that jet
launching can be produced from this interaction region with physical
temperatures and densities similar to those of stellar chromospheres
(Goodson et al. 1997, von Rekowski \& Brandenburg 2004, 2006).
Jet launching from the interface between the magnetic rotor (the star) and the
disc is much more efficient than a pure disc wind, firstly because the
centrifugal gear is higher closer to the star and secondly because of the heavy
mass load onto
the field lines at this point. Thus, to conclude this series, we have computed
the radiative output  from numerical simulations of outflow launching from star-
disc interaction. It will be shown that the radial flow expansion, i.e. the
outflow from the disc-star interaction region, is able to reproduce
the observed broadenings for magnetospheric fields of about 1-2 kG.
In Section~2, the characteristics
of the simulations from von~Rekowski \& Brandenburg (2004, 2006,
hereafter vRB04 and vRB06, respectively), that are used in this work,
are summarised. In Section~3, the procedure to derive
line fluxes and profiles from the simulations
is described. The analysis of the output profiles is developed in Section~4.
A discussion on the implications and relevance of the results
is deferred to Section~5, as well as the main conclusions.

\section{The T Tauri Star-disc system}

\subsection{The magnetospheric star-disc model}

The interaction between the star and the Inner disc Boundary Layer\footnote{
The shearing layer between the region dominated by the stellar magnetic
field and the gravitationally dominated Keplerian disc has a finite extent.}
generates a
sheared region fed by turbulent,
magnetised material from the accretion disc. Shear amplifies the stellar
magnetic field,
producing a strong toroidal magnetic field component.
This toroidal field and the associated magnetic pressure push the stellar
poloidal field
away from the stellar/disc rotation axis, inflating and opening  the poloidal
field lines
in a {\it butterfly like pattern} thus producing a current layer
between the stellar and the disc dominated regions as displayed in
Fig.~1. The magnetic link between the
star and the disc is broken and re-established periodically by
magnetic reconnection. The opening angle of the current layer,
as well as its extent, depends on the stellar and disc fields,
the accretion rate and the ratio between
the inner disc radius and the stellar rotation frequency.

In the magnetospheric models of vRB04, a slow, hot and dense outflow driven mostly 
by poloidal magnetic pressure is  emanating from stellar regions close to the rotation
axis while fast, cooler and less dense magneto-centrifugally accelerated outflows 
are emanating from lower stellar latitudes and from the inner disc.
The main difference with the magnetospheric models of Goodson et al. (1997) is
that in the Goodson models the axial (stellar) jet
is fast and well-collimated - and driven by magneto-centrifugal processes (see
also Hirose et al. 1997) -
and the inner disc wind is divergent, whereas in vRB04 the stellar wind is slow
but on the other hand
some collimation is clearly seen in the inner (magneto-centrifugally
accelerated) disc wind.
However, the stellar wind seen in Romanova et al. (2002) is also slow,
travelling into a rarefied corona.
A very fast (warm and dense) stellar wind is seen in stellar dynamo models of
vRB06; this wind is mostly
driven by toroidal and poloidal magnetic pressure, as well as gas pressure.

\begin{figure}
\includegraphics[width=80mm]{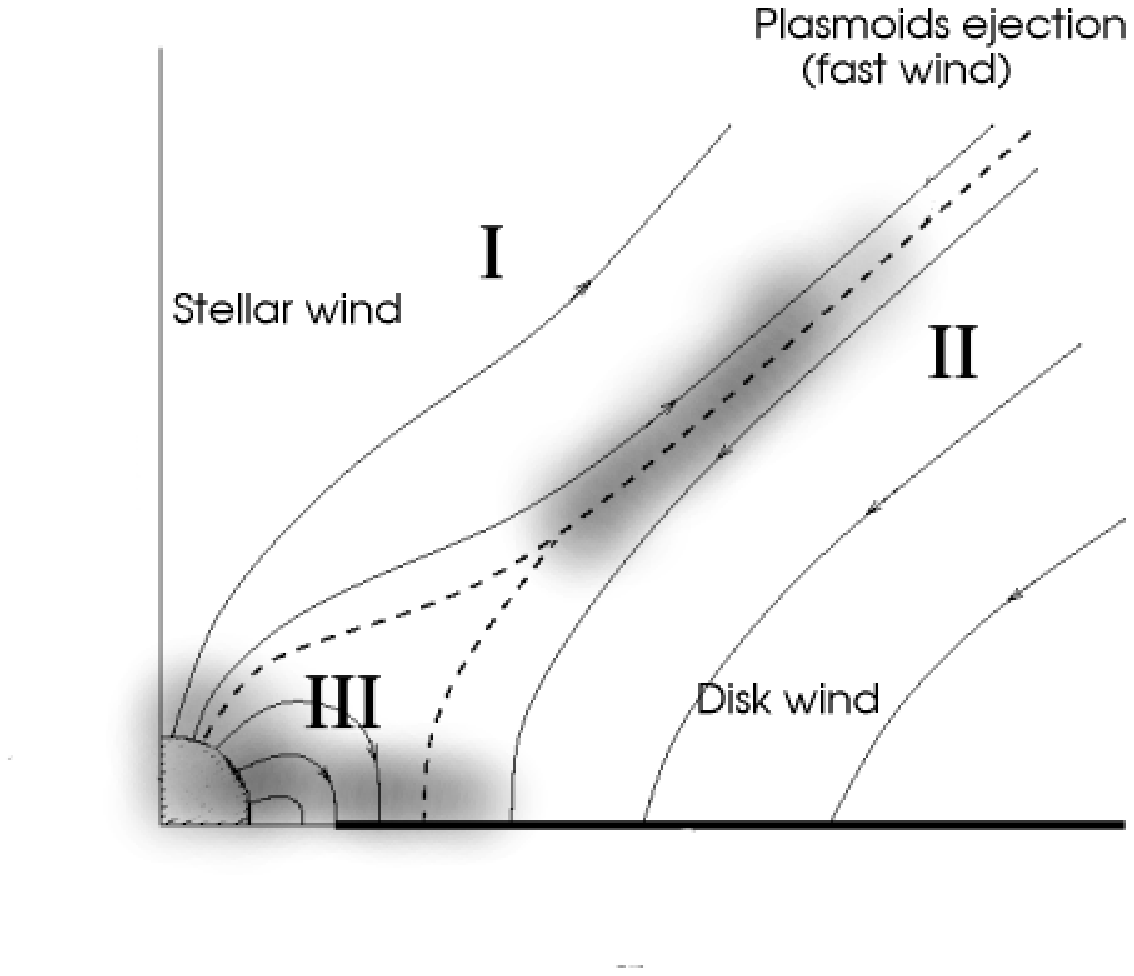}
 \caption{The interaction between the stellar magnetic field and the
disc twists the stellar field lines due to the differential rotation.
The toroidal magnetic field generated out of the poloidal flux and
the associated  magnetic pressure tends to push the poloidal field lines
outwards, inflating them, and eventually braking the magnetic link between the
star and the disc (boundary between regions I and II). Three basic regions can
be defined:
Region I dominated by the stellar wind, Region II dominated by
the disc wind and Region III dominated by stellar magnetospheric
phenomena. The dashed lines trace the boundaries between these
three regions. The continuous lines indicate the topology
of the magnetic field and the shadowed areas represent regions where
magnetic reconnection events are likely to occur, producing
high energy radiation and particles (from G\'omez de Castro 2004).}
\end{figure}

\subsection{Basic characteristics of the numerical models}

Though a full description of the numerical simulations can be found in
vRB04 and vRB06, we should briefly summarise their main properties.
In both papers, the evolution of the flow, magnetic field, density and
temperature
is found by solving the continuity, the Navier--Stokes and the mean-field
induction MHD equations
for an axisymmetric system in cylindrical polar coordinates, assuming a
piecewise polytropic model.
Dynamo action in the disc (if present) is prescribed by a standard
$\alpha^2 \Omega$ dynamo (e.g. Krause \& Raedler 1980), where
$\alpha$ is the mean-field $\alpha$-effect and $\Omega$ is the angular
velocity of the orbiting gas. $\alpha$-quenching is included
so that the disc dynamo saturates at a level close to equipartition
between magnetic and thermal energies. The code uses
dimensionless variables that have been scaled using as
reference values a typical sound speed of the coronal gas (100 km s$^{-1}$)
and a typical surface density at the surface of the disc (1 g cm$^{-2}$).
Furthermore, the mass of the star has been taken as 1 M$_{\odot}$ (solar mass),
the mean specific weight as $\mu = 0.6$ and the polytropic index as $\gamma = 5/3$.
The computations have been carried out in a domain
of extent 0.2~AU in the radial ($\varpi$) direction
and $\pm 0.1$~AU above/below the disc mid-plane;
mesh sizes are $\delta \varpi = \delta z = 0.001$AU.
The inner edge of the disc is at 4 stellar radii (in the models with stellar
dipolar magnetosphere; cf.\ Table~1) or at 2.4 stellar radii (in the model with
stellar dynamo), which in both cases corresponds to 12 solar radii.
The disc extends to the outer boundary of the computational domain.

The vRB04 models represent a step forward over the self-similar
warm disc winds models of Paper~II. Around the inner edge of the disc, the interaction
with the stellar magnetosphere and the stellar wind is taken into account.
Also, the rigidity of the self-similarity constraint
is lost. Furthermore, the Lorentz force $\vec j \times \vec B$
is included in the Navier-Stokes equation, as well as the generation
of magnetic fields by the standard $\alpha ^2 \Omega$ dynamo in the
disc. The saturation level of the disc dynamo is
governed by the equipartition between magnetic and thermal energies
in the disc. In accordance with observations of protostellar star-disc systems, 
the model implements a dense, cool disc embedded in a rarefied, hot disc
corona. This is achieved by prescribing an entropy contrast between the disc and
corona so that specific entropy is smaller within the disc and larger in the
corona, and by choosing hydrostatic equilibrium as the initial state. Therefore,
detailed modelling of the coronal heating physics is avoided
by prescribing a "thermal" frontier between disc and corona
that is based on the observational properties of accretion discs
(piecewise adiabatic model in disc, corona and star).
However, temperature is evolving with time, in particular in the disc
corona as one can clearly see in the relevant figures. Assuming a perfect gas,
the continuity equation is re-formulated as an evolution equation for specific
enthalpy, which is directly related to temperature. The specific enthalpy is
then combined with the gravitational potential to form the potential enthalpy.

In a sense, the disc in the vRB04 simulations can be understood as a
dynamical boundary that regulates the extra electromotive force caused by
the disc dynamo and taps the mass flow into the wind and onto the accreting
star. vRB04 adds
to the dynamics of the self-similar models in Papers~I \& II but
shares a similar thermal treatment allowing a straightforward
comparison.

The full set of simulations in vRB04 and vRB06 explores the role of disc-star
interaction in  mass ejection for various configurations of the stellar field (a
dipolar magnetosphere or a mean-field $\alpha^2 \Omega$ dynamo generated field) 
and strengths (models with different stellar surface fields are analysed). 
The simulations also explore the role of magnetically active discs (with $\alpha^2 
\Omega$ dynamos) in the dynamical evolution of the system
by comparing the active disc evolution with that of  passive discs.
In all the simulations, the three basic winds: hot stellar wind, warm disc
wind and episodic ejections from reconnection in the star-disc current
layer are readily identified. In this work
the radiative output has been calculated only for a subset of these simulations.

Firstly, we have selected a reference model to analyse whether and how the
temporal evolution of the simulated star-disc system shows in the profiles of
the spectral tracers of the outflows. This is model "M1" in vRB04.
The stellar magnetosphere is initially a dipole threading the disc but with time
can freely evolve outside the anchoring region away from the stellar
surface; the dipole 
is aligned with the disc axis and has a strength at the stellar surface of
$B_* \sim 1$kG. The disc is assumed to be undergoing
a significant dynamo action with $\alpha _0 = -0.1$. This is the so-called {\it
"Reference"} model in Table~1. The thermal properties of the disc wind at the
launching points in the corona are similar to those derived in Paper~II
for warm disc winds, i.e.:
\begin{eqnarray}
T &= &6.4865\times 10^4 {\rm K }y^{0.05}\left\langle \mu \right\rangle
\left( \frac{M_{*}}{M_{\odot }}\right) \left( \frac{\varpi _{0}}{0.1{\rm AU}}
\right)^{-1}
\end{eqnarray}
where $M_{*}$ is the mass of the T Tauri star, $\varpi _0$ is the radial
distance of the wind foot point (in the disc corona)
and $y$ is the ratio
between the outflow density at $\varpi _0$ and at the Alfv\'en radius
(see Paper II for more details). The temperature drops as $\varpi ^{-1}$
at the base of the disc wind, similarly to what derived in vRB04
for the {\it Reference} model (see
i.e. Fig.~2 in vRB04). Notice that in Paper~II these conditions are set
to reproduce the observed large scale properties of protostellar jets
following Vlahakis \& Tsinganos (1999).
In vRB04, the thermal properties of the wind are
derived from the specific enthalpy, $h$, that is included in the modified
Navier-Stokes and continuity equations (von Rekowski et al. 2003).
The system
is piecewise adiabatic and the equation of state a polytrope, thus the
sound speed is given as $c_s ^2 = (\gamma -1)h$, with $\gamma = 5/3$,
and the temperature is derived directly from $h = c_{\rm p} T$.

In addition,  we have selected a set of models to study the variation of the
quiescent profiles with the strength of the magnetospheric field.
Models with  $B_* \sim 2$kG and  $B_* \sim 5$kG have been analysed. These
models are quoted as  {\it "Mag-2kG"} and {\it "Mag-5kG"}, respectively.
As in the reference
model, the disc is assumed to be undergoing dynamo action with $\alpha _0 = -
0.1$.

The next step has been to explore whether the presence of an active or
passive disc can be detected in the profiles. For this purpose we have
analysed a simulation similar to that of the {\it "Reference"} model but with a
passive disc (no disc dynamo);
this model is quoted as {\it "Passive"} in Table~1.

Finally, we have analysed how the profiles change with the characteristics of
the
stellar field, i.e. we have investigated whether introducing an active dynamo in
the star modifies
the line profiles. This model is quoted as {\it "Dynamo*"}; the star is
considered to
be fully convective with the stellar dynamo rooted up to about 5 solar radii and
$\alpha_0 = +1.5$
for the stellar dynamo.

A summary with the names and main magnetic properties of the models is given in
Table~1.

\begin{table}
 \centering
  \caption{Summary of the simulations analysed}
  \begin{tabular}{lll}
  \hline
   Iden & Stellar field & disc field \\
  \hline
{\it Reference} & Dipolar, B$_* \sim$ 1kG & $\alpha ^2 \Omega$ dynamo with
$\alpha_0= -0.1$ \\
{\it Mag-2kG}& Dipolar, B$_* \sim$ 2kG & $\alpha ^2 \Omega$ dynamo with
$\alpha_0= -0.1$ \\
{\it Mag-5kG}& Dipolar, B$_* \sim$ 5kG & $\alpha ^2 \Omega$ dynamo with
$\alpha_0= -0.1$ \\
{\it Passive}& Dipolar, B$_* \sim$ 1kG  & No dynamo \\
{\it Dynamo*} & $\alpha ^2 \Omega$ dynamo, $\alpha_0= 1.5 $& $\alpha ^2 \Omega$
dynamo with $\alpha_0= -0.15$ \\
   \hline

\end{tabular}
\end{table}

The vRB04 models analyse the dynamical evolution of the star-disc interaction during
total run times between $\sim 150$~d and $\sim 1900$~days. However, the time
scale analysed in this work is significantly shorter:
about $75$~d in the {\it "Reference"} model where the effect of
outburst compared to quiescence is studied in the profiles.
This shorter time scale is, in general, about twice the typical matter
crossing time from the outer to the inner disc border: $(\varpi _{\rm out}
-\varpi _{\rm in})/u_{\varpi} {\bf \approx} (0.2-0.06){\rm AU}/7 {\rm km} {\rm s}^{-1}
\sim 34.5$~d. In vRB04, matter is injected locally wherever and
whenever it is needed in the disc region, in order to model mass accretion from 
the outer parts of the disc to the inner parts. 
This might produce artificially high mass loss rates in the
long term, especially around the inner disc edge where mass outflow
is more efficient (see vRB04 for a detailed discussion on the implications
of this approach).  Selecting periods shorter
than the crossing time minimises the effect of matter injection.

\begin{figure}
\includegraphics[width=80mm]{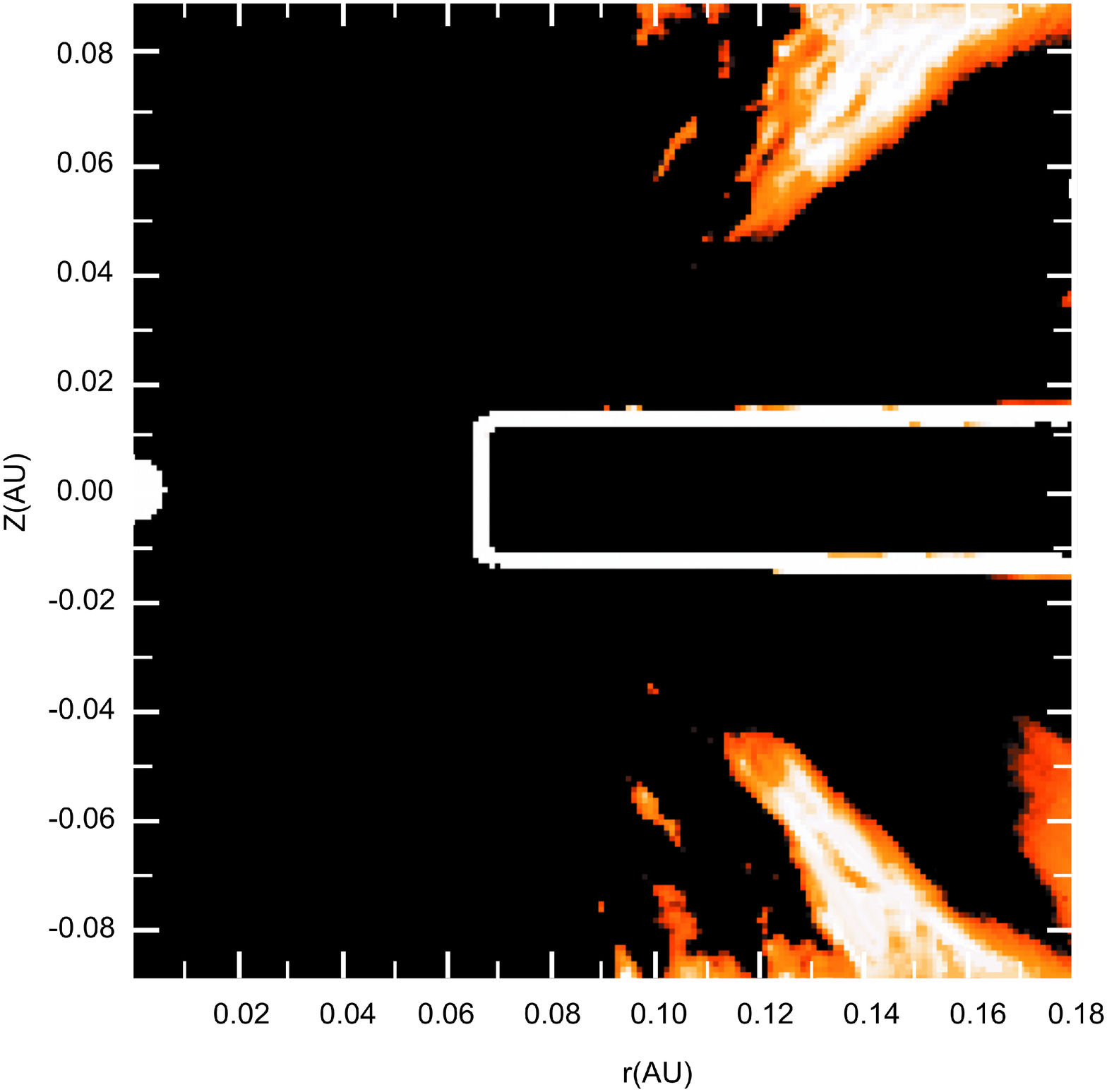}
\includegraphics[width=80mm]{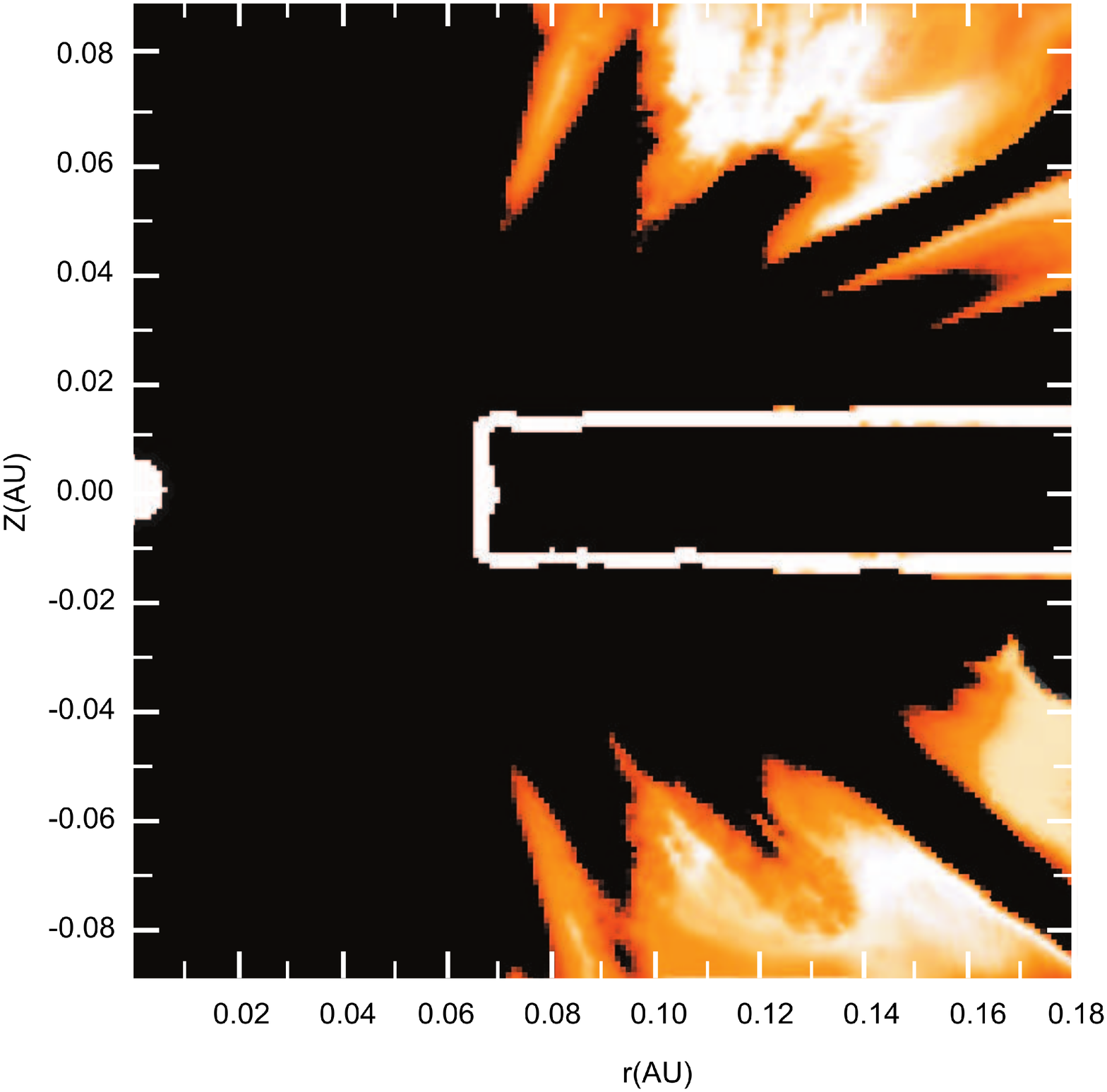}
\caption{Si~III] emissivity maps of the reference model during quiescence (top)
and in outburst (bottom). Notice that at outburst the Si~III] emitting region
is large and reaches closer to the star. }
\end{figure}
\section{Computation of the line profiles}

Flow kinematics is handy described
by three components:
\begin{itemize}
\item Motion along the rotation (or $z$) axis (vertical flow component): $V_z$
\item Rotation given by the toroidal flow component: $V_{\rm t}$
\item Expansion away from the rotation axis (radial flow component):
$V_{\varpi}$
\end{itemize}
The numerical simulations provide as output the density $\rho_k$,  temperature
$T_k$,
the three velocity components ($V_{z,k}$, $V_{\rm t,k}$, V$_{\varpi,k}$) and the
three
magnetic field components at any given cell, $k$, of the computational mesh.
As the models are axisymmetric, the simulations provide the same temperature
and density for a cylindrical ring around the disc axis. As the C~III]$_{1908}$
and the Si~III]$_{1892}$ lines are semiforbidden, the output radiation
can be directly computed from the emissivity and the volume of the
ring. The spectral distribution of the radiation is then computed for a given
$k$-ring using
the projection of the ring velocity into the line of sight. At any point in the
ring, the velocity  projected into the line of sight\footnote{The minus sign has
been
set to indicate the velocity shift as seen by the observer, so
negative velocities (or ``blueshifts'') are associated with gas
motions towards the observer.} is given by
\begin{equation}
V_{{\rm rad}, \alpha}  = - V_{\varpi,k} \cos \alpha \sin i -
V_{z,k} \cos i+V_{\rm t,k} \sin \alpha \sin i,
\end{equation}
where $i$ is the inclination of the star (the angle between the line
of sight and the disc axis) and $\alpha$ is the angle between the
stellar meridian that contains the line of sight and the meridian where the gas
parcel under consideration in the $k$-ring is located;
each ring has been subdivided into
100 cells with $\Delta \alpha = 2 \pi /100$ for the calculation
of the profile. The temperature $T_k$ is taken into account to compute the
effect of thermal broadening.

	The emissivity of any $k$-ring is calculated for the semiforbidden
line of Si~III]. C~III] emissivity was also calculated but the density is
high and the line is collisionally quenched in most of the simulated area
thus it was not further studied. The atomic parameters
for the lines were taken from the Chianti Atomic Data
Base\footnote{URL: www.damtp.cam.ac.uk/user/astro/chianti/}
(see Appendix~A). Collisional rates have been calculated using thermally
averaged collision strengths making use of Burgess \& Tully (1992) fits
and rules. For each $T_k$,
the ionisation fraction has been calculated and the electron density, $n_e$,
has been derived. The electron temperature, $T_e$, is assumed to be equal
to $T_k$ in each given $k$-ring.

Si~III] emissivity, $j$, corresponding to the transition
between levels j and m is given by
\begin{eqnarray}
j = & n_j A_{j,m} h\nu _{j,m} =  \nonumber \\
& f(n_e,T_e) \frac {n_j}{n_{Si III}}\frac {n_{Si III}}{n_{Si}}
\frac {n_{Si}}{n_H}\frac {n_H}{n} n A_{j,m} h\nu _{j,m}.
\end{eqnarray}
The population of the relevant levels ($ {n_j}/{n_{Si III}}$) has been
computed directly with the Chianti code for each temperature and electron
density. The ionisation fractions (${n_{Si III}}/{n_{Si}}$) have been calculated
assuming ionisation equilibrium. Solar abundances are used for ${n_{Si}}/{n_H}$ and
the ionisation fractions to determine ${n_H}/{n}$. $ f(n_e,T_e)$ is a correction
factor that takes into account the collisional quenching of the line\footnote{$$
f(n_{\rm e},T_{\rm e}) = \frac {A_{j,m}}{A_{j,m}+n_{\rm e} q_{j,m}},
$$
where $n_{\rm e} q_{j,m}$ and $ A_{j,m} $ are the rates for downward
collisional and radiative transitions, respectively. Note that
when the line is collisionally quenched, then
$f(n_{\rm e},T_{\rm e}) \propto n_{\rm e}^{-1}T_{\rm e}^{1/2}< \Upsilon >
(T_{\rm e}) $.}.

For a given inclination, the theoretical Si~III] profiles
are computed from the simulation data by adding the contributions from all the
$k$-rings with
$\varpi \geq 4 R_*$; this condition was partly set to avoid confusion
with any stellar-associated component. In addition, the numerical
simulations analysed (vRB04, vRB06) are not adequate to follow the thermal
instabilities
in the boundary layer between the star and the disc and, as a consequence, the
physical conditions
and the derived radiative output are not reliable for this region of the
computational mesh.

For each simulation in
Table~1, the Si~III] profile has been computed for each time unit (approximately
3 days)
to track the profile variability, to separate quiescence from outburst and
to compute the line flickering during the quiescence phase.

\section{Analysis of the output profiles}

A given profile is a set of pairs ($V_{rad}$, $f$)  where $f$ is
the total flux radiated by the outflow into the line of sight
with projected radial velocity in the range [$V_{rad}-5$~km~s$^{-1}$,
$V_{rad}+5$~km~s$^{-1}$]. $V_{rad}$ ranges from -300~km~s$^{-1}$ to
300~km~s$^{-1}$, taken in steps of 10~km~s$^{-1}$ that oversample the thermal
broadening (typically $\sim 30$~km~s$^{-1}$) by a factor of 3.

The system alternates episodic ejections of matter (outbursts) with quiescent
periods where accretion is dominant (see vRB04).  During outburst, the magnetic
star-disc link is basically broken: the stellar dipolar field lines connecting to
the disc are opening up and matter is flowing along the open stellar and disc
field lines forming an outer stellar wind and enhanced inner disc wind.
In Fig.~2, Si~III] emissivity maps are shown of the reference model at
four different times: the first three images are during the quiescence period
while the last one is during outburst.

A reliable comparison between the theoretical profiles and the observed profiles
requires the identification of a set of statistically meaningful quantities to
describe the profiles. We have used for this purpose:

\begin{itemize}
\item The {\it velocity of the peak of the line}, $V_p$.
\item The {\it  line asymmetry}, $A$, defined as:
$
A = \frac{ \Sigma _{v_b} ^{0} f}{ \Sigma _{v_b} ^{v_r} f}
$, where $f$ represents the flux at projected radial velocity $v$,
$v_b$ marks the minimum radial velocity of the blue wing of the line,
and $v_r$ marks the maximum radial velocity of the redwing of the line.
\item The {\it line centroid}, $<V>$, defined as:
$
<V> = \frac { \Sigma _{v_b} ^{v_r} f * v} { \Sigma _{v_b} ^{v_r} f}.
$
\item The {\it dispersion}, $\sigma$, defined as:
$
\sigma = \frac { \Sigma _{v_b} ^{v_r} f * (v^2 -<V>^2)} { \Sigma _{v_b}
^{v_r} f}.
$
\end{itemize}
Moreover, as the redwing of the profile may have a variable contribution
from mass infall (see RY~Tau profiles in GdCV07),
dispersion and
line centroid have also been evaluated only for the blue wing of the profile:

\begin{itemize}
\item The {\it line centroid of the blueshifted emission}, $<V_b>$, defined as:
$
<V_b > = \frac { \Sigma _{v_b} ^{0} f * v} { \Sigma _{v_b} ^{0} f}.
$
\item The {\it dispersion in the blueshifted emission}, $\sigma_b$, defined
as:
$
\sigma _b = \frac { \Sigma _{v_b} ^{0} f * (v^2 -<V_b>^2)} { \Sigma _{v_b}
^{0} f}.
$

\end{itemize}
\begin{figure*}
\includegraphics[width=160mm]{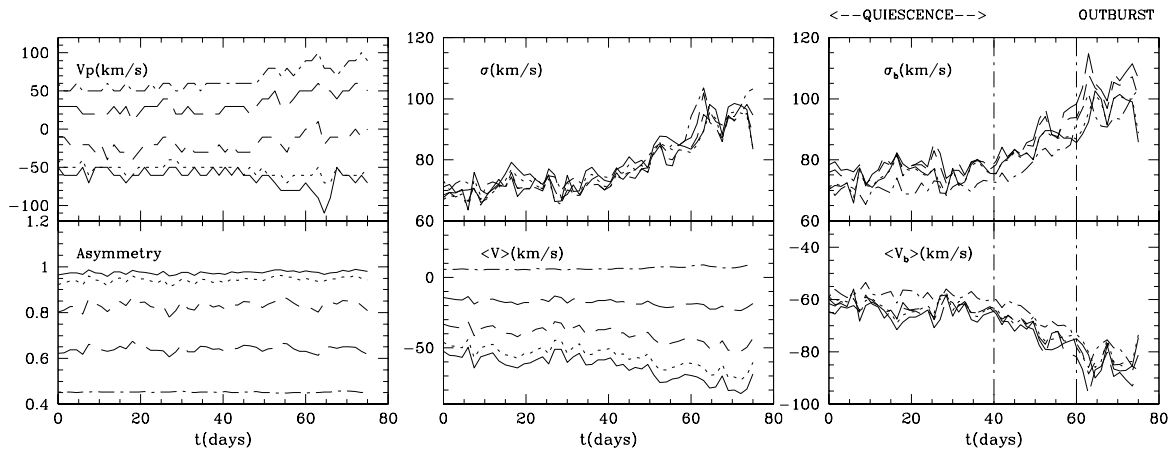}
 \caption{Temporal variations of the statistical quantities used to characterise
the line profiles
in the simulation of the (magnetospheric) reference model. The data
corresponding to
inclinations of 0$^0$, 23$^0$, 45$^0$, 68$^0$ and 90$^0$ are represented by
solid, dotted, short-dashed, long-dashed and dot-dashed lines, respectively.
The time refers to after the start of the analysed time series.}
\end{figure*}

In Fig.~3 the temporal variations of these quantities are plotted for a whole
quiescence-outburst
cycle for the {\it "Reference"} model. The quiescence period can be readily
identified,
as well as the outburst period when the profiles are more weighted to the
blue and broader than in quiescence. In other words, the quantities more
sensitive
to trace mass ejection are: $\sigma$, $\sigma _b$ and $<V _b>$ which, in turn,
are
rather insensitive to the inclination of the source with respect to the line of
sight.

The effects in the Si~III] profiles due to an outburst are shown in Fig.~4 for
the reference model: the line broadens and the emission region expands towards
the boundaries of the computational domain. However, during outburst there is
also some contribution from matter infall (accretion) to the profile (see bottom
panels in Fig.~4), when some matter is channelled along the rising and closing
field lines into an infalling accretion flow - at the same time as enhanced
winds are starting or ceasing to form.
In summary, a clear correlation between outflow and accretion signatures is
predicted from these numerical simulations.

Though these results are extremely promising because the correlation between
infall and outflow signatures has been reported for various cTTSs
(i.e. RU~Lup by Stempels \& Piskunov 2002, RW~Aur by G\'omez de 
Castro \& Verdugo 2003 and AA~Tau by Bouvier et al 2007), 
they should be considered just as a first order approach; the axisymmetry
assumption imposed in the code and in the calculation of the radiative output,
forces one to assume that large plasma tori are ejected (accreted) along
the disc axis at the time of outburst (as the smoke rings produced by
master smokers). However, in the T Tauri systems mass ejection occurs in the
form of blobs (plasmoids) from reconnecting magnetic fields, which have a much
smaller emitting
volume than the tori assumed in this axisymmetric computation. As a consequence
the fluxes predicted in this paper for outburst (both infall and outflow) are
overestimated.
For this reason, we should concentrate on the analysis of the profiles
during the quiescence phase.

\begin{figure}
\includegraphics[width=80mm]{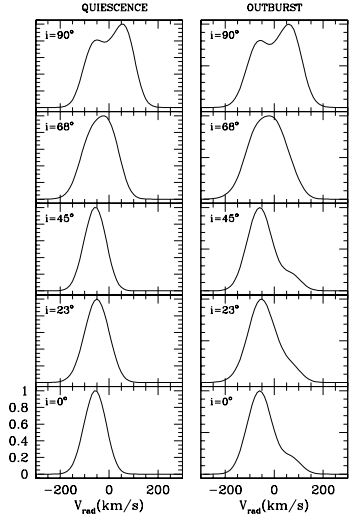}
 \caption{Theoretical profiles of the Si~III] line derived from the
reference model of the numerical simulations. The time averaged profile in
quiescence
is plotted in the left hand panels and the time averaged profile in outburst
is plotted in the right hand panels, for various inclinations.}
\end{figure}

During quiescence, flickering produces variations
of $\sigma_b$ and $<V_b>$ that are highly correlated, as shown in Fig.~5 for
two numerical simulations ({\it Reference} and {\it Dynamo*})
and two possible inclinations; the correlation coefficient
is between 0.96 and 0.9998 depending upon the
simulation. This high correlation indicates that the shape of the line profile
does not vary significantly - to the first moment - during quiescence, and, in
turn, the variations in the profile caused by mass ejections are averaged out by
the large broadening induced by the thermal motions of the gas. The largest
variations
in $\sigma_b$ are observed for the pole-on (inclination 0$^o$) orientation.

The time-average values $\tilde{\sigma} _b$ and $<\tilde{V}_b>$ are given for
each simulation
and five inclinations during the quiescence period in Table~2.
The dispersion about these average values is
also given; note that this dispersion reflects the short-time scale
variability (the flickering) of the highly non-stationary star-disc interaction.
The theoretical line flux calculated from the simulations is provided in the
last column. As no
extinction effects are taken into account in the calculation of the line
profiles and fluxes, the fluxes given in the table are the same for all the
orientations in each model\footnote{The circumstellar extinction is inclination
dependent since it is caused by dust lifted from the disc}. The flickering
during quiescence produces slight variations in the line fluxes in the range of
$\sim 10$\%.
The fluxes derived are similar to those derived for warm disc models (see
Table~2 in Paper~II).
Average profiles for the various models and three inclinations are displayed in
Fig.~6.

\begin{figure}
\includegraphics[width=80mm]{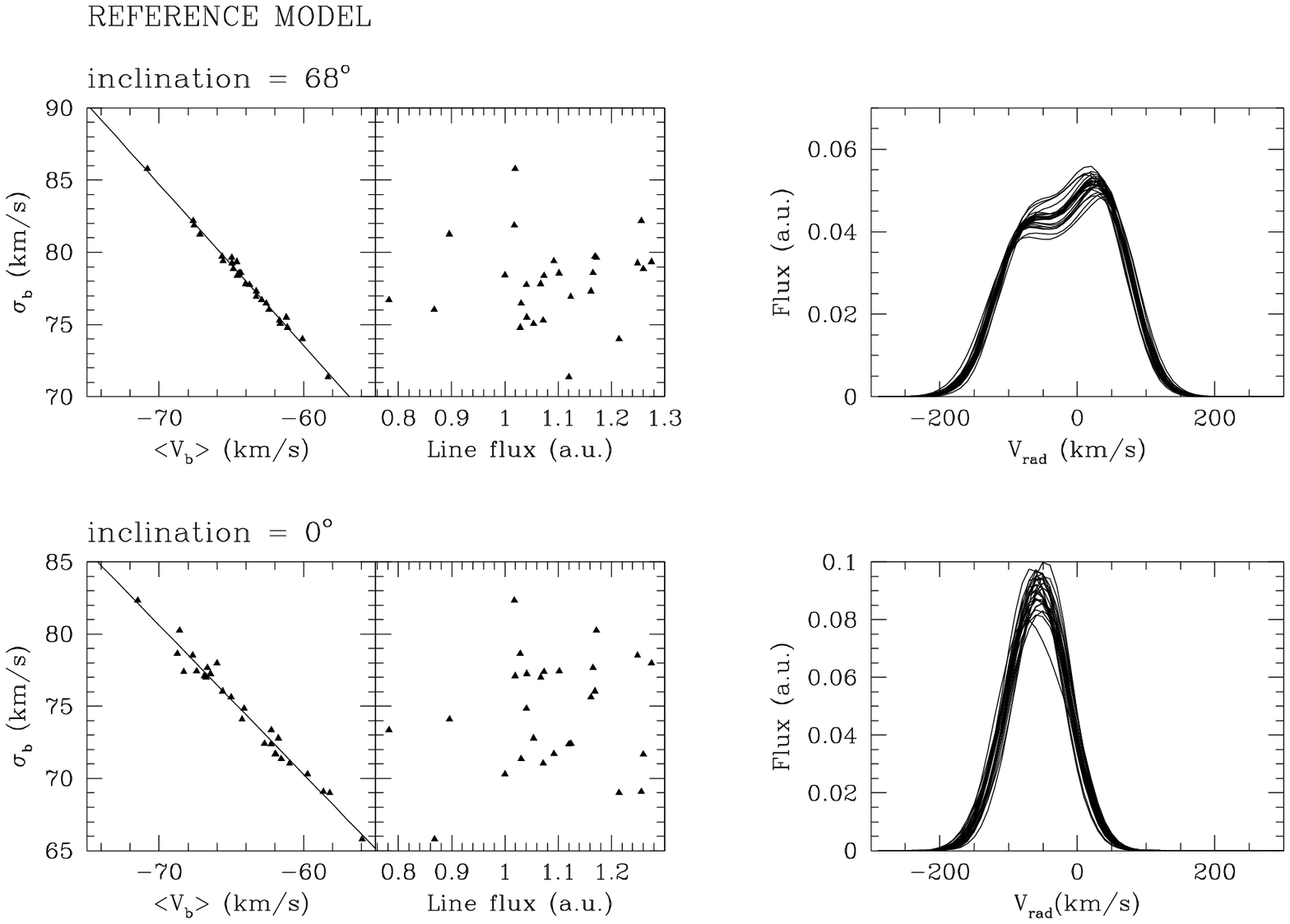} \label{fig5a}
\includegraphics[width=80mm]{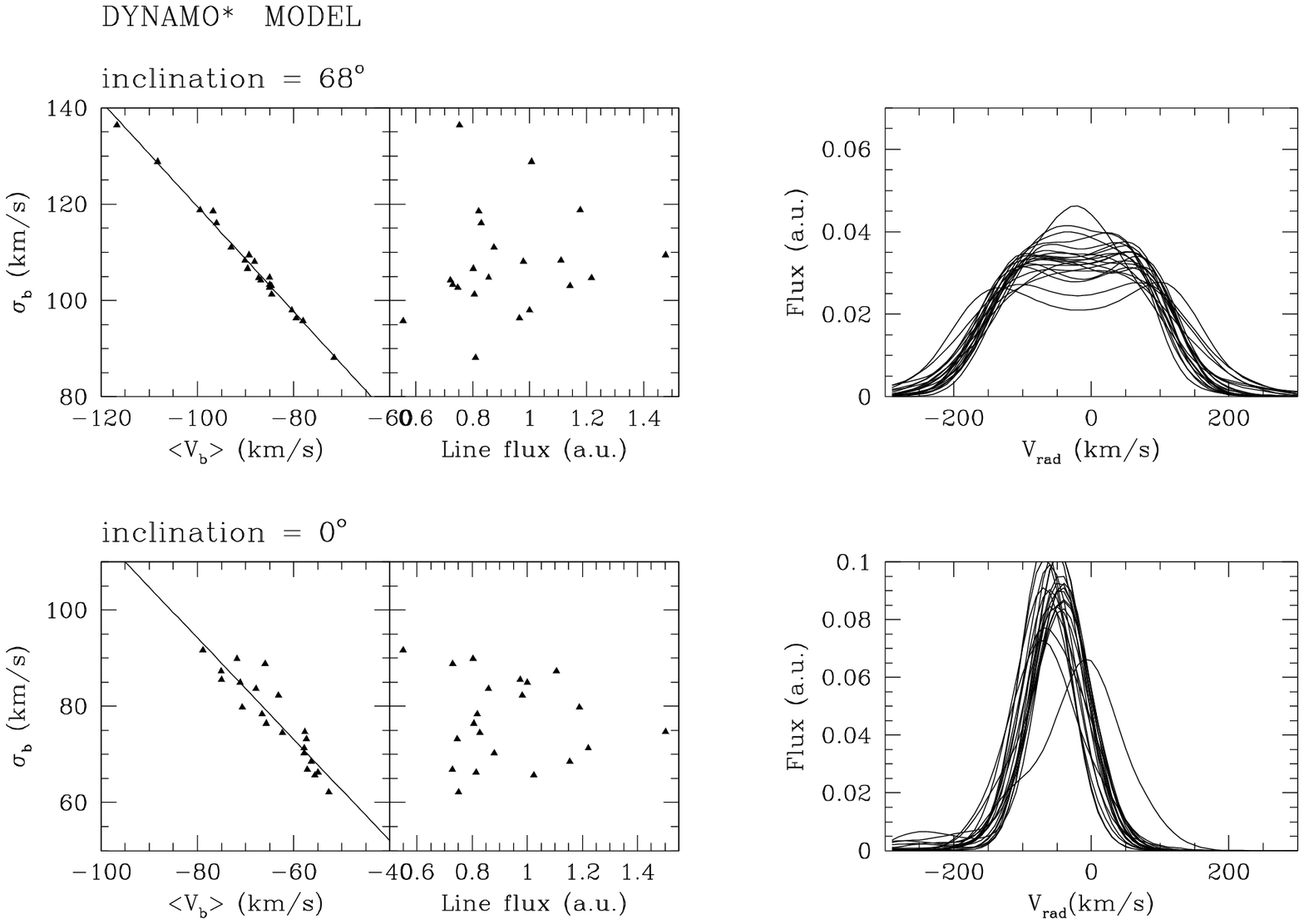} \label{fig5b}
 \caption{Left panels: variation of $\sigma _b$  versus $<V_b>$ and versus the
line
flux (in arbitrary units) during quiescence  and inclinations
68$^o$ and 0$^o$ (in the top and bottom panels, respectively). The temporal line
profile variations
during quiescence are represented in the right panels.
{\it Top:} {\it "Reference model"}, {\it bottom:} {\it "Dynamo*"} model.
}
\end{figure}

\begin{table}
 \centering
  \caption{The $\tilde{\sigma} _b$ and $<\tilde{V}_b>$ quantities (averaged over
time in the quiescence
  period) derived from the simulations}
  \begin{tabular}{@{}lllll@{}}
  \hline
   Iden & Inclination & $<\tilde{V}_b>$  & $\tilde{\sigma} _b$ & Flux \\
	&		&		& 		& (erg~s$^{-1}$~cm$^{-2}$) \\
  \hline
{\it Reference} &  0$^o$ & -64 $\pm$ 4 & 75 $\pm$ 4 & $(4.0 \pm 0.5)\times 10^{-
8}$\\
	    & 23$^o$ & -63 $\pm$ 3 & 74 $\pm$ 3 & \\
            & 45$^o$ & -63 $\pm$ 2 & 77 $\pm$ 3 & \\
            & 68$^o$ & -64 $\pm$ 3 & 78 $\pm$ 3 & \\
	    & 90$^o$ & -58 $\pm$ 2 & 71 $\pm$ 2 & \\
{\it Mag-2kG}& 0$^o$ & -81 $\pm$ 7 & 93 $\pm$ 8 & $(3.9 \pm 0.6)\times 10^{-8}$
\\
	    & 23$^o$ & -81 $\pm$ 6 & 96 $\pm$ 7 & \\
            & 45$^o$ & -85 $\pm$ 7 & 104 $\pm$ 8 & \\
            & 68$^o$ & -90 $\pm$ 8 & 107 $\pm$ 9 & \\
	    & 90$^o$ & -82 $\pm$ 8 & 97 $\pm$ 9 & \\
{\it Mag-5kG}& 0$^o$ & -89 $\pm$ 14 & 103 $\pm$ 16 & $(3.9 \pm 0.6)\times 10^{-
8}$ \\
	    & 23$^o$ & -90 $\pm$ 12 & 109 $\pm$ 15 &  \\
            & 45$^o$ & -101 $\pm$ 11 & 124 $\pm$ 14 & \\
            & 68$^o$ & -108 $\pm$ 11 & 130 $\pm$ 13 & \\
	    & 90$^o$ & -101 $\pm$ 14 & 121 $\pm$ 18 & \\
{\it Passive}&  0$^o$ & -45 $\pm$ 6 & 54 $\pm$ 6 & $(1.9 \pm 0.2)\times 10^{-8}$
\\
	    & 23$^o$ & -45 $\pm$ 6 & 57 $\pm$ 6 & \\
            & 45$^o$ & -48 $\pm$ 6 & 63 $\pm$ 7 & \\
            & 68$^o$ & -50 $\pm$ 6 & 65 $\pm$ 7 & \\
	    & 90$^o$ & -46 $\pm$ 4 & 60 $\pm$ 5 & \\
{\it Dynamo*} &  0$^o$ & -64 $\pm$ 8 & 77 $\pm$ 9 & $(3.1 \pm 0.8)\times 10^{-
8}$\\
	    & 23$^o$ & -70 $\pm$ 5 & 86 $\pm$ 6 & \\
            & 45$^o$ & -83 $\pm$ 7 & 102 $\pm$ 8 & \\
            & 68$^o$ & -89 $\pm$ 10 & 108 $\pm$ 11 & \\
	    & 90$^o$ & -85 $\pm$ 10 & 103 $\pm$ 11 & \\   \hline

\end{tabular}
\end{table}

\begin{figure}
\includegraphics[width=80mm]{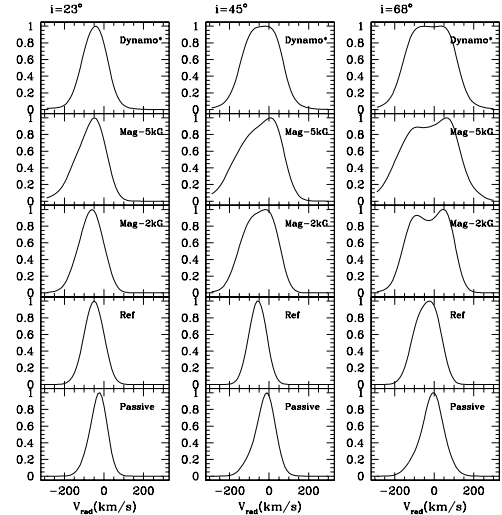}
 \caption{Theoretical profiles of the Si~III] line during quiescence for
the five numerical models analysed (see Table~1) and three inclinations with
respect to the line of sight: 23$^o$, 45$^o$ and 68$^o$.}
\end{figure}

Several general conclusions can be derived from the analysis of the quantities
and profiles:

\begin{itemize}

\item
Enhancing the dipolar stellar magnetic field enhances $\tilde \sigma _b$,
$<\tilde V _b>$ and their respective dispersions. Thus these quantities
are good tracers of the magnetic field strength and therefore of the wind
launching capability.
The reason is that the centrifugal gear,
for a given accretion rate, depends on the location
of the Alfv\'en surface. Stronger fields push this surface further away from the
star, thereby  raising the wind terminal speed as well as increasing the radius
$\varpi$
of the ring radiating in Si~III], and thus increasing also the dispersion of the
projected
velocity in the region where the emitting gas is located.

\item
$\tilde \sigma _b$ and $<\tilde V _b>$ are sensitive to the viewing angle (the
inclination
of the line of sight with respect to the star) for strong stellar dipolar fields
or for
stellar fields that are dynamo induced. In general, the larger the inclination
the
larger the values of $\tilde \sigma _b$ and $<\tilde V _b>$; however,
the maxima of $\tilde \sigma _b$ and $<\tilde V _b>$ are achieved at
inclination $\sim 68 ^o$, which roughly corresponds to the angle between the
field lines and the disc axis in the main ejection region:
the current sheet region between the disc field and the stellar field.

\item
Passive discs produce narrow lines. There are two
reasons for that. As discussed in vRB04, the initial magnetic field
in the disc due
to the penetrating stellar magnetosphere is expelled out of the disc and the
wind flows
nearly parallel to the disc surface in the low corona. As a result,
the velocity dispersion is smaller. In addition, the Alfv\'en
surface is located at the innermost boundary of the star-disc system and
thus the centrifugal gear is negligible.

\item
Active stellar dynamos are able to generate a rapid outflow even for a
moderately strong
magnetic field.
The reason is because in the {\it "Dynamo *"} model the stellar wind is mostly
driven by both
poloidal and toroidal magnetic pressure, as well as by gas pressure.

\end{itemize}

\section{Comparison with observations and Summary}

In Fig.~7, the time-averaged values $\tilde \sigma _b$ and $< \tilde V _b>$
are plotted for the various models studied in this work.
The error bars mark the dispersion around the average values
in quiescence, caused by the flickering. All the inclinations
are plotted and models are colour coded. The general
trends described in the previous section are clearly apparent.

\begin{figure}
\includegraphics[width=80mm]{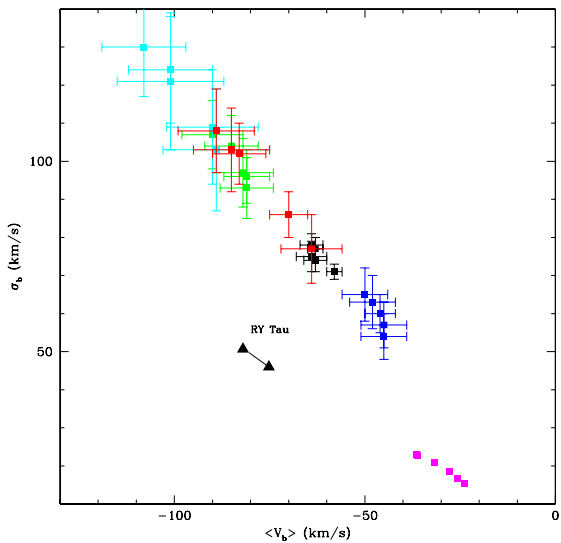}
 \caption{Representation of $\sigma_b$ versus $<V_b>$ for all models and
inclinations
of Table~2. The time-averaged values $\tilde \sigma_b$ and $<\tilde V_b>$ are
represented as squares, while the error bars indicate true temporal variations
of the profiles
during quiescence. Dark blue, black, green and light blue are used to represent
the models: Passive, Reference, Mag-2kG and Mag-5kG, respectively. Red is used
to represent
the Dynamo* model. Purple squares at the bottom right represent predictions from
the
warm disc wind models by G\'omez de Castro \& Ferro-Font\'an (2005). The {\it
observed
values} for the TTS: RY~Tau are plotted as two big black triangles, representing
the
two extreme values that can be derived from the observations carried out with
the HST in 1998
(G\'omez de Castro \& Verdugo 2001, 2007).}
\end{figure}

$\tilde \sigma _b$ and $< \tilde V _b>$ are also plotted for
warm disc winds (purple squares) for the analytical solutions
of G\'omez de Castro \& Ferro-Font\'an (2005). As expected,
warm disc wind solutions are less efficient than winds
driven from the star-disc boundary layer.

There is a correlation between $\tilde \sigma _b$ and $<\tilde V _b>$,
extending from the profiles derived in the numerical simulations
of this work to warm disc winds (to some extent). This correlation 
traces the increasing efficiency of MHD centrifugally driven winds
and shows that larger terminal speeds are associated with larger bulk motions
at the base.  Notice that inclination effects are difficult to sort out in this
diagram.

\subsection{The observed Si~III] profiles}

The information available on the Si~III] profiles of TTSs come from the
observations carried out by G\'omez de Castro et al (2003, hereafter 
GdCVFF03) with the HST. The Si~III] profiles are plotted in Fig.~8, as well as 
the C~III] lines needed for the density diagnostic. 
All the targets (unless AK~Sco) have reported jets allowing a clear 
identification of the outflow signature (see Table~3).
As the size of the spectrograph slit was 0.2 arcseconds,
any extended structure within 28 AUs could contribute to the line flux.

The contribution from the large scale jet to the line flux
can be discriminated by the density of the emitting plasma. 
The electron density in the jet decreases very 
rapidly with $z$: from $\sim 10^9 - 10^{10}$~cm$^{-3}$ at the base, 
to $\sim 10^4- 10^6$~cm$^{-3}$ in the jet due to 
the rapid radial expansion of MHD centrifugally driven winds.
At the low densities of the jet, the forbidden lines of [Si~III]($\lambda 1983$) and
[C~III]($\lambda 1907$) are stronger than their semiforbidden
counterparts. Also, the ratio
SiIII]/CIII] is $<1$ (Keenan et al 1992), as observed in the
UV spectra of protostellar jets and Herbig-Haro objects 
(see G\'omez de Castro \& Robles 1999 for a compilation).

This is not observed in the HST/STIS spectra of TTSs as shown in Fig.~8
(from GdCVFF03). Typically, the Si~III] and C~III]
lines have similar strengths. The forbidden 
[C~III] line is only detected in T~Tau and RU~Lup and, only in
T~Tau, has a non negligible contribution to the Si~III] flux. Thus 
Si~III] emission is produced at the base of the outflow in
most of the sources.

\begin{table}
 \centering
  \caption{Properties$^{a}$
 of the TTSs observed in Si~III]}
  \begin{tabular}{llll}
  \hline
   Iden & Jet velocity & $V \sin i$ & i \\
        & (km/s) & (km/s) & ($^o$)\\
  \hline
DE Tau & -125 $^{(1)}$   & 10 $^{(4)}$   &  - \\
T Tau  & -111 $^{(2)}$   & 20.1 $^{(4)}$ & 20 $^{(6)}$ \\
RW Aur & -167 $^{(2)}$   & 17.2 $^{(4)}$ & 40 $^{(6)}$ \\
RU Lup & $<-170>$ $^{(3)}$ &    -     & 24 $^{(7)}$ \\
AK Sco &    -       & 19 $^{(5)}$   & 63 $^{(5)}$ \\
RY Tau & -80 $^{(2)}$    & 52.2 $^{(4)}$ & 86 $^{(8)}$\\
   \hline

\end{tabular}
\begin{tabular}{l}
References: (1) Hartigan et al 1995;
(2) Hirth et al 1997; \\(3) Takami et al 2001; (4) Clark \& Bouvier 2000;\\
(5) Andersen et al 1989; (6) Ardila et al 2002; \\(7) Stempels et al 2007;
(8) Muzzerolle e al 2003. \\
\end{tabular}
\end{table}

\begin{figure}
\includegraphics[width=80mm]{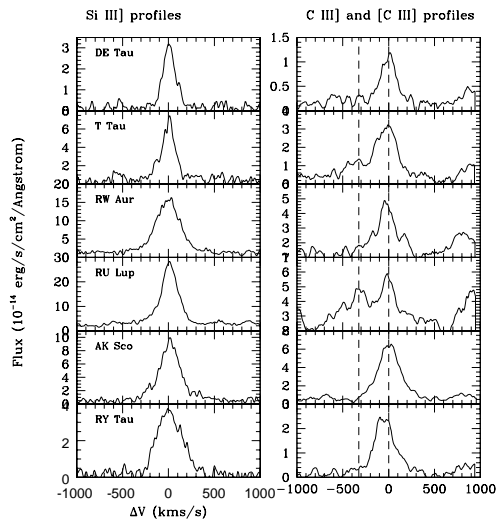}
 \caption{Measured Si~III], C~III] and [C~III] profiles of TTSs (from GdCVFF03). 
The rest velocity corresponds to the rest wavelength of the 
Si~III] and C~III] lines in the left and right panels, respectively.
The "velocities" of the [C~III] and C~III] lines are marked with dashed lines
in the right panel for clarity.} 
\end{figure}

The Si~III] flux may include additional contributions from 
accretion shocks and the stellar magnetosphere/atmosphere. 
Discriminating between these three components is much more difficult
since they share similar temperature and density regimes.

RY~Tau is the only TTS that has been observed twice in the Si~III] lines
allowing to separate cleanly the contribution from accretion and outflow
becoming an ideal test for the models. $\tilde \sigma_b$ and $<\tilde V_b>$ 
for RY~Tau (from GdCV07) are also plotted in Fig.~7. 
Notice that the line centroid is too blueshifted for the observed dispersion.
This is because the contribution from the outflow is dominated by 
a narrow blueshifted component with a long tail to redwards shifted
velocities (see Fig~1 in GdCV07). 
This disagreement between theory and observations can be caused by the
lack of cylindrical symmetry of the outflow (i.e. mass ejected 
as blobs from the current layer).

\subsection{The $\tilde \sigma$ versus $<\tilde V>$}

The $\tilde \sigma$ versus $<\tilde V>$ diagram is 
better suited to break the inclination degeneracy and
examine the relevance of outflows with respect to magnetospheric
processes in TTSs. The centroid of the line emission from the wind
moves to higher blueshifts as the inclination with respect to the line of sight
decreases however, no significant inclination effects are expected 
from magnetospheric models. 

The values of $\tilde \sigma$ and $<\tilde V>$ derived from
the Si~III] profiles (see Fig.~8) are plotted for comparison.
Notice that $<\tilde V> \simeq 0$ for most sources and
that there is not any trend related  with the inclination.
This suggests that that Si~III] emission is dominated by circumstellar
structures rather than by disc winds. In fact, RW~Aur, the
star with the highest dispersion, is surrounded by a torus/belt of ionized plasma
(G\'omez de Castro \& Verdugo, 2003). However, even stars that have not
this kind of circumstellar structures, as DE~Tau, display profiles
with large broadenings that cannot be accounted solely by stellar rotation.
Thus, it seems that the dominant contribution to the Si~III] emission is
produced in extended magnetospheric structures where the
large broadening traces either the accretion
flow (Hartmann 2009) or with large macro-turbulence fields,
as the hypothesised to reproduce the
broad profiles of AK~Sco (G\'omez de Castro 2009).

A small contribution from the outflow to the Si~III] flux
is detected RU~Lup, T~Tau and RY~Tau (once the contribution
from accretion is subtracted). It shows as a bluewards shift 
of the line centroid. Unfortunately, further observations 
are required to disentangle magnetospheric and wind 
contributions  prior to an accurate diagnostic of the wind drives.

We would like to conclude this work with a word of caution. This research is
limited by the
characteristics of the simulations analysed, such as
the axisymmetry,
the rather small extent of the computational domain and
the very efficient  tapping of thermal energy imposed by the polytropic approach
in the simulations.
More realistic models implementing radiative cooling and adaptive mesh
refinement
techniques to track the thermal instabilities are currently under development.
These new simulations will produce more detailed models of the dynamics
and the radiative output from the disc-star interface; however, they will not
modify the strong dependence of the profiles on the inclination
shown in Fig.~8, which is caused by the coupling between ejection and rotation
and holds even on scales of 0.1~AU.

\begin{figure}
\includegraphics[width=80mm]{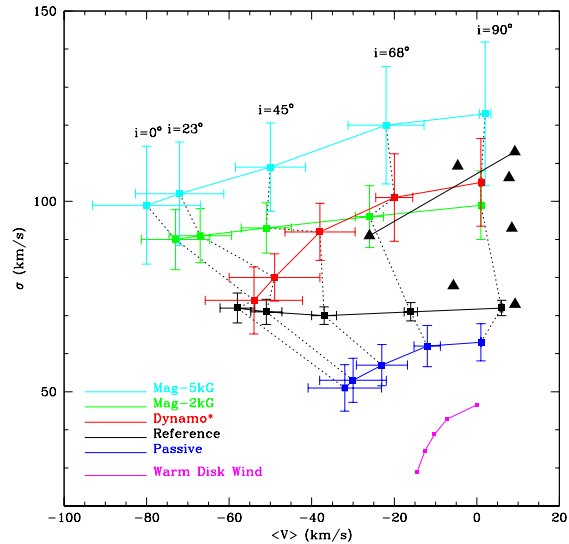}
 \caption{Representation of $\sigma$ versus $<V>$ for all models and
inclinations
of Table~2. The time-averaged values $\tilde \sigma$ and $<\tilde V>$ are
represented as squares, while the error bars indicate true temporal variations
of the profiles
during quiescence. Dark blue, black, green and light blue are used to represent
the models: Passive, Reference, Mag-2kG and Mag-5kG, respectively. Red is used
to represent
the Dynamo* model. Purple squares at the bottom right represent predictions from
the
warm disc wind models by G\'omez de Castro \& Ferro-Font\'an (2005). The {\it
observed
values} for all the TTSs observed with the HST/STIS (DE~Tau, AK~Sco, RY~Tau,
RW~Aur, T~Tau and
RU~Lup) are plotted as big black triangles. The two observations of RY~Tau are
plotted; the less blueshifted include the contribution from the accretion flow
as in the observing campaign in 2001.}
\end{figure}

\section*{Acknowledgments}
Ana I. G\'omez de Castro acknowledges the support from the Ministries of Science
and
Education (MEC) and Science and Innovation (MiCIn) of Spain for support through
research grants AYA2007-67726, AYA2008-06423 as well as the regional government
of Madrid for support through grant S-505/ESP/0237.

\appendix
\section{Atomic parameters for the calculation of the line emissivities}

\begin{table*}
\begin{flushleft}
\caption[]{Atomic parameters of the semiforbidden lines}
\begin{tabular}{llccccccccccc}

\hline
Element & Wavelength & $j$& $m$ & A$_{j,m}$ & w$_j$ & Type$^{(1)}$ & C$^{(2)}$ &
\multicolumn{5}{c}{$< \Upsilon _{j,m} > ^{(3)} $ } \\
  \cline{9-13}  \\
& $\lambda _{j,m}(\AA)$ & & & s$^{-1}$ &  & & & 0.0 & 0.25 & 0.50 & 0.75 & 1.00
\\
&     & & & & & & & & & \\ \hline
Si III &1892.033 & 3 & 1 & 1.544$\times 10^{4}$ & 3 & 2& 0.4 & 2.931 & 1.655 &
1.063 & 0.4002 & 0 \\
C III & 1908.737 & 3 & 1 & 97.32   & 3 & 2& 1.1 &0.3688 & 0.3419 & 0.3047 &
0.2102 & 0.0888\\
\hline
\end{tabular}
\begin{tabular}{ll}
$^1$ & Transition type according to Burgess \& Tully (1992) classification.Type
2
includes transitions which are induced by\\
     &  either an electric multipole or a magnetic multipole, e.g. forbidden
transitions.\\
$^{2,3}$ & Collisional rates have been computed following Burgess \& Tully
(1992).
   Thermally averaged collision strengths,\\
  & $ < \Upsilon _{j,m} > (x)$, are given by interpolation
for each temperature, $T_{\rm e}$, using the  coefficients  indicated in the
table with \\
& $x (T_{\rm e}) = (k_B T_{\rm e}/E_{j,m})/(C+ (k_B T_{\rm e}/E_{j,m}))$ and
$E_{j,m} = h c /\lambda_{j,m}$.
The collision rate is: $8.6233 \times 10 ^{-6} n_{\rm e}  T_{\rm e}^{-1/2} <
\Upsilon _{j,m}> /w_j$. \\
& ($k_B$: Boltzman constant; $h$: Planck constant; $c$: speed of light;
$w_j$: statistical weight of level $j$.) \\

\end{tabular}
\end{flushleft}
\end{table*}

\label{lastpage}

\end{document}